\documentclass[preprint,12pt,amsmath,amssymb]{elsarticle}

 \usepackage{epsfig}
 \usepackage{bm}
 \usepackage{graphicx}
 \usepackage{subcaption} 
\DeclareUnicodeCharacter{2061}{}
\usepackage{amssymb}

\biboptions{sort&compress}

\journal{Optics Communications}

\begin{document}

\begin{frontmatter}



\title{Generation of robust temporal soliton trains by the multiple-temporal-compression (MTC) method}


\author[label1]{Andr\'e C. A. Siqueira}
 \ead{andrechaves.physics@gmail.com}
\affiliation[label1]{
            organization={Departamento de Física, Universidade Federal de Pernambuco},
            city={Recife},
            postcode={50670-901}, 
            state={Pernambuco},
            country={Brazil}}

\author[label2]{Guillermo Palacios}

\affiliation[label2]{organization={Comissão Nacional de Energia Nuclear, Centro Regional de Ciências Nucleares do Nordeste -- CNEN/CRCN-NE},
            city={Recife},
            postcode={50740-545}, 
            state={Pernambuco},
            country={Brazil}}

\author[label3]{Albert S. Reyna}

\affiliation[label3]{organization={Programa de Pós-Graduação em Engenharia Física, Unidade Acadêmica do Cabo de Santo Agostinho, Universidade Federal Rural de Pernambuco},
            city={Cabo de Santo Agostinho},
            postcode={54518-430}, 
            state={Pernambuco},
            country={Brazil}}

\author[label4,label5]{Boris A. Malomed}

\affiliation[label4]{organization={Department of Physical Electronics, School of Electrical Engineering, Faculty of Engineering},
            postcode={69978}, 
            state={Tel Aviv},
            country={Israel}}

\affiliation[label5]{organization={Instituto de Alta Investigación, Universidad de Tarapacá},
            city={Casilla 7D},
            state={Arica},
            country={Chile}}

\author[label1]{Edilson L. Falc\~ao-Filho}
\author[label1]{Cid B. de Ara\'ujo}

\begin{abstract}
We report results of systematic numerical analysis for multiple soliton generation by means of the recently reported multiple temporal compression (MTC) method, and compare its efficiency with conventional methods based on the use of photonic crystal fibers (PCFs) and fused silica waveguides (FSWs).  The results show that the MTC method is more efficient to control the soliton fission, giving rise to a larger number of fundamental solitons with high powers, that remain nearly constant over long propagation distances. The high efficiency of the MTC method is demonstrated, in particular, in terms of multiple soliton collisions and the Newton’s-cradle phenomenology.
\end{abstract}


\end{frontmatter}
\section{\label{sec:level1}Introduction}

In the course of the last decades, the investigation of temporal solitons has been widely developed for the study of various ultrafast optical phenomena, from the viewpoint of both fundamental research and applications \cite{sysoliatin2007soliton,tai1988fission,driben2013newton,braud2016solitonization,gordon1986theory,mitschke1986discovery,xiang2011controllable,skryabin2003soliton,knight1996all,birks1997endlessly,russell2003photonic,liu2012all,dudley2002numerical,hult2007fourth,agrawal2013nonlinear,dudley2006supercontinuum}. Temporal solitons have been studied in the context of optical communications, where they may be employed as data bits \cite{hasegawa1973transmission, mollenauer1988demonstration, hasegawa1995solitons, agrawal2012fiber, Iannone1998nonlinear, agrawal2001applications, mollenauer2006solitons}. The generation of self-confined solitary waves is commonly achieved by balancing the group-velocity dispersion (GVD) and self-phase modulation (SPM) in nonlinear (NL) media. For this purpose, the power of the incident beam must exceed the self-confinement critical level ($P_{crit}$), so that the nonlinearity allows the input pulse to maintain its temporal envelope during the propagation \cite{agrawal2013nonlinear}. Thus, if the incident power is higher than $P_{crit}$, the input pulse evolves into a fundamental soliton with the remaining energy shed off as dispersive waves. If the incident power is much higher than $P_{crit}$, multiple fundamental solitons may be generated, under specific conditions \cite{antikainen2012phase, erkintalo2010giant, husakou2001supercontinuum, bose2016implications, arteaga2018soliton, bose2016study, bose2018dispersive, driben2010solitary, zhao2022effects, skryabin2010colloquium}. 

One widely used method for generating temporal solitons is based on the dispersion and nonlinearity engineering in photonic crystal fibers (PCFs) \cite{knight1996all, birks1997endlessly, russell2003photonic, liu2012all}. This methodology has been useful for investigating the generation of temporal solitons in the near-infrared and visible ranges, due to the ability of PCFs to shift the anomalous dispersion region towards much lower wavelengths than those attainable with standard fibers. It is relevant to stress that, due to the high nonlinearity, low confinement loss and tunable chromatic dispersion that PCFs exhibit, they have been commonly used to investigate the evolution of fundamental and higher-order solitons. The first-order (fundamental) solitons are the only ones that strictly maintain a constant profile during the propagation, while higher-order solitons show a periodically varying shape. Under the action of perturbations, higher-order solitons split into several fundamental solitons, each centered on a different wavelength, in the process called soliton fission \cite{agrawal2013nonlinear}. Although soliton
fission works well in PCF, its efficiency in producing multiple-soliton trains is poor. This occurs because during the fission strong intrapulse Raman scattering (IRS) is induced as a result of significant temporal compression around the central region of the input pulse, caused by the generation of the first soliton exhibiting a very broad supercontinuum spectrum.The dissipative effect of the IRS limits the soliton propagation over long distances, and reduces the number of generated solitons, as a large portion of the energy remains confined in the first soliton.

Recently, we reported a new methodology, called multiple temporal compression (MTC), to generate multiple temporal ultrashort solitons in a more efficient way \cite{siqueira2023generation, siqueira2022generation}. The MTC method is based on a specific version of the general scheme of the nonlinearity management \cite{malomed2006soliton}, viz., the propagation of a laser pulse through a composite medium made up of alternating self-focusing and self-defocusing segments, both with normal group-velocity dispersion. During propagation in the first segment, the input pulse accumulates positive linear ($\phi_{disp}$) and NL ($\phi_{nl}$) phases due to the GVD and self-focusing effects, respectively. Consequently, the pulse is positively chirped. Subsequently, a partial compensation of $\phi_{nl}$ occurs, due to the self-defocusing effect, when the pulse propagates in the second segment. This compensation is not complete under to the accumulation of $\phi_{disp}$ that occurs in both segments due to the action of the normal dispersion. Thus, instead of the compensation of the positive chirp in the second segment in the course of the propagation, it generates pairs of temporal solitons that are temporally symmetric with respect to the pulse’s center. Furthermore, in the course of the longer propagation, more temporal solitons are generated from the leading and trailing edges, and eventually from the central region of the pulse.

In this work, we numerically solve the generalized NL Schr\"{o}dinger equation to compare the efficiency in generating multiple temporal solitons by using the MTC method \cite{siqueira2023generation, siqueira2022generation} and previous methods based on the use of photonic crystal fibers (PCFs) and fused silica waveguides (FSWs). The efficiency of the methods is compared according to the number of temporal solitons formed from the same input conditions, as well as considering the energy distribution in the multiple-soliton train during the fission and the ability of each soliton to maintain its properties unchanged in the course of the subsequent propagation.

\section{\label{sec:level2}Numerical analysis of the multiple temporal solitons process}

The generation and propagation of temporal solitons is studied in the framework of the generalized NL Schr\"{o}dinger equation (NLSE), written as \cite{dudley2002numerical, hult2007fourth, agrawal2013nonlinear, dudley2006supercontinuum}: 


\begin{eqnarray} 
 \frac{\partial A}{\partial z} - \left(\sum\limits_{n\geqslant 2}\beta_{n}\frac{i^{n+1}}{n!}\frac{\partial^{n} A}{\partial T^{n}}\right) = i\gamma_{0}\left(1 + \frac{i}{\omega_{0}} \frac{\partial}{\partial T}\right)( \left(1- f_{R}\right)A|A|^{2} + \hspace*{1.01 cm} \nonumber \\ + f_{R}A\int_{0}^{\infty}h_{R}(\tau)|A(z,T- \tau)|^{2}d\tau),
\label{eq:1}
\end{eqnarray}
where the term in the parentheses on the left-side of the Eq. (1) describes the high-order dispersion, while the right-side of the Eq. (1) describes the third-order NL optical effects with $\gamma_{0}= \omega_{0}n_{2}(\omega_{0})/cA_{eff}$ being the third-order NL coefficient. The temporal derivative in the first parenthesized factor on the right-hand side represents effects such as the self-steepening and optical shock formation. In the second factor, the fractional contribution of the delayed Raman response to the NL polarization is represented by $f_R$, with the first term representing the temporal SPM and instantaneous Raman contribution. The second term represents the convolution integral describing the delayed Raman response, which leads to effects such as the IRS and the soliton self-frequency shift.

To perform the numerical solutions of Eq. (1), we use the fourth-order Runge-Kutta in the interaction picture (RK4IP) method, which was reported to be more accurate compared to other methods such as the conventional split-step Fourier method \cite{hult2007fourth}. The simulations were started for input pulses with the hyperbolic secant shape, pulse duration time $T_{FWHM} = 90 fs$, and the input peak power varying from $P_0 = 2.5 kW$ to $30 kW$. For the purpose of the comparison the central pulse’s wavelength ($\lambda_0$) is chosen in such a way that the materials available for the realization of the MTC, PCF and FSW methods exhibit close absolute values of the GVDs with opposite signs. The NL parameters of the medium used in each procedure are described below.

The generation and propagation of temporal solitons using the PCF, which implies the pulse propagation in the single medium (with length $L =150 mm$), was considered in the framework of the well-established model \cite{dudley2002numerical, hult2007fourth}. In this case, for the central pulse wavelength $850 nm$, we take GVD coefficients in Eq. (1) as $\beta_{2} = -1.27\cdot10^{-2} ps^{2}m^{-1}$ (the anomalous-dispersion regime), $\beta_{3} = 8.11\cdot10^{-5} ps^{3}m^{-1}, \beta_{4} = -1.32\cdot10^{-7} ps^{4}m^{-1}, \beta_{5} = 3.03 \cdot10^{-10}ps^{5}m^{-1}, \beta_{6} = - 4.19\cdot10^{-13} ps^{6}m^{-1},$ and $\beta_{7} = 2.57\cdot10^{-16} ps^{7}m^{-1}$. 

The MTC method was also analyzed considering the same central pulse wavelength, $\lambda_0 = 850 nm$, but in the normal-dispersion regime, by using fused silica as the matrix for the two segments of the NL medium. By using pure fused silica in the first segment, a positive NL refractive index is obtained, with the GVD coefficients $\beta_{2} = + 3.22\cdot10^{-2}ps^{2}m^{-1}, \beta_{3} = 2.94\cdot10^{-5}ps^{3}m^{-1}, \beta_{4} = -1.67\cdot10^{-8}ps^{4}m^{-1}, \beta_{5} = 4.61\cdot10^{-11}ps^{5}m^{-1}, \beta_{6} = - 1.28\cdot10^{-13}ps^{6}m^{-1},$ and $\beta_{7} = 4.31\cdot10^{-16}ps^{7}m^{-1}$, which were calculated from the derivation of the Sellmeier expression \cite{malitson1965interspecimen}. On the other hand, the second segment of the NL medium includes metal nanoparticles in fused silica to achieve the self-defocusing behavior \cite{bose2016implications, arteaga2018soliton, bose2016study, bose2018dispersive, driben2010solitary, zhao2022effects, zhang2017nonlinear, reyna2017high, kassab2018metal, reyna2022beyond}, while maintaining the  medium in the normal-dispersion regime. Another strategy may use birefringent NL crystals, such as LiNbO$_3$, BBO, and KTP, in the second segment of the NL medium, as negative second-order cascade effects are strong enough to compensate for excessive positive Kerr nonlinearity. In both cases, the second segment with an effective defocusing nonlinearity, $n_{2,eff} < 0$, and normal dispersion is used \cite{desalvo1992self, ashihara2002soliton, bache2008limits, guo2014few, vsuminas2017second, conforti2013extreme}.

For completeness, we also consider the temporal soliton generation procedure that uses a $150 mm$ length FSW. In this case, it was not possible to obtain a normal-dispersion regime. However, for comparison with the MTC method, the central pulse wavelength of $1585 nm$ is chosen for use in the framework of the FSW method so that the absolute value of GVD is the same considered in both methods. Thus, from the derivation of the Sellmeier expression, for fused silica at $1585 nm$, we obtain: $\beta_{2} = -3.22\cdot10^{-2}ps^{2}m^{-1}, \beta_{3} = 16.56\cdot10^{-5}ps^{3}m^{-1}, \beta_{4} = - 5.63\cdot10^{-7}ps^{4}m^{-1}, \beta_{5} = 2.73\cdot10^{-9}ps^{5}m^{-1}, \beta_{6} = - 1.61\cdot10^{-11}ps^{6}m^{-1},$ and $\beta_{7} = 1.11\cdot10^{-13}ps^{7}m^{-1}$ \cite{malitson1965interspecimen}. 

For correct comparison between the three methods, $|\gamma_{0}|= 0.045 W^{-1} m^{-1}$ is adopted as the magnitude of the NL coefficient of all involved media, which corresponds to the reference value reported for a PCF 
\cite{dudley2002numerical, hult2007fourth}. In the PCF- and FSW-based methods, one has $\gamma_{0} > 0$ and the NL medium length $L = 150 mm$. However, in the framework of the MTC method, we use the first segment of length $10 mm$ with the NL coefficient $\gamma_0^{(1)} = |\gamma_{0}|$, and the second segment of length $140 mm$ with $\gamma_0^{(2)} = -|\gamma_{0}|$ . 

The Raman parameters used to model the influence of the IRS were obtained from Refs. \cite{dudley2002numerical, hult2007fourth, agrawal2013nonlinear}, as all the NL media considered in this work are based on fused silica. Thus, the fractional contribution of the delayed Raman response to the nonlinear polarization is taken as $f_R = 0.18$, and the Raman response function is $h_R(\tau) = \tau_1 (\tau_1^{-2} + \tau_2^{-2}) exp(-\tau/\tau_2)sin(\tau/\tau_1)$, where $\tau_1 = 12.2 fs$ and $\tau_2 = 32.0 fs$.

\section{\label{sec:level3}Results and discussions}

Figure 1 shows the generation and propagation of solitons produced by the temporal evolution of the pulse intensity, as obtained from simulations of Eq. (1) adjusted to modeling the PCF and FSW methods. Figures 1(a) and 1(c), obtained for input peak powers $5 kW$, show that both methods generate the first soliton after the incident pulse has traveled the distance $\approx 20 mm$ ($ \approx 25 mm$ for PCF, once we have $|\beta_2^{(PCF)}|<|\beta_2^{(FSW)}|$  ). It is clearly seen that the first soliton carries a large part of the input beam power, with the slight excessive power carried by dispersive waves ejected around the soliton. For higher input powers, the energy remaining after the creation of the first soliton drives the generation of secondary solitons, i.e., the higher-order soliton breaks up into fundamental solitons through fission. For instance, Figs. 1(b) and 1(d) show that, respectively, five or three temporal solitons are generated by the PCF or FSW method, for input power $25 kW$. Although the PCF method has been shown to be more efficient than the FSW, both demonstrate low peak powers of the secondary solitons, as well as power loss in the course of propagation under the action of the strong IRS.

\begin{figure*}
    \begin{subfigure}[b]{0.50\textwidth}
        \includegraphics[width=\textwidth]{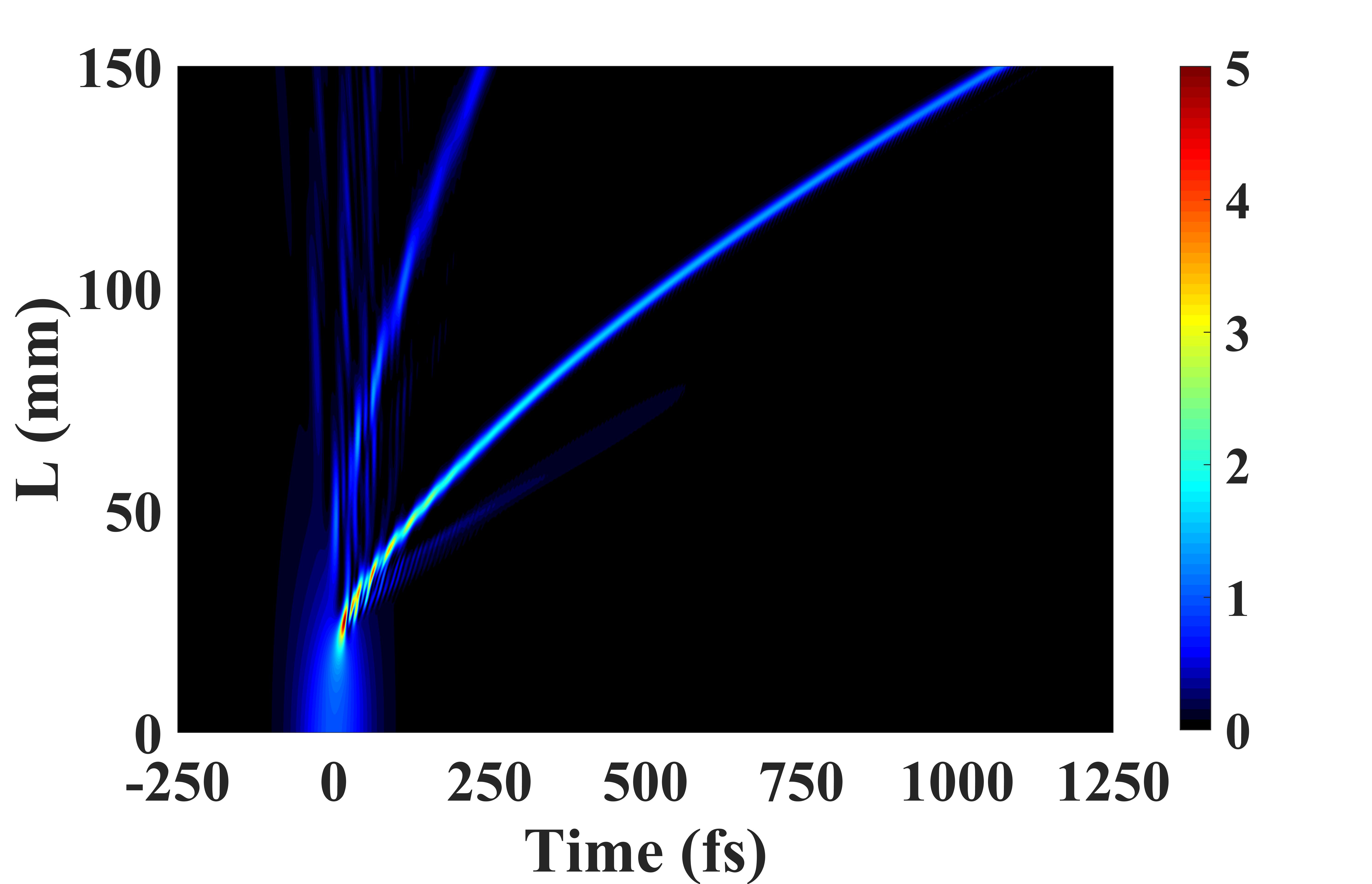}
         \caption{}
        \label{fig:1a}
    \end{subfigure}
    \begin{subfigure}[b]{0.50\textwidth}
        \includegraphics[width=\textwidth]{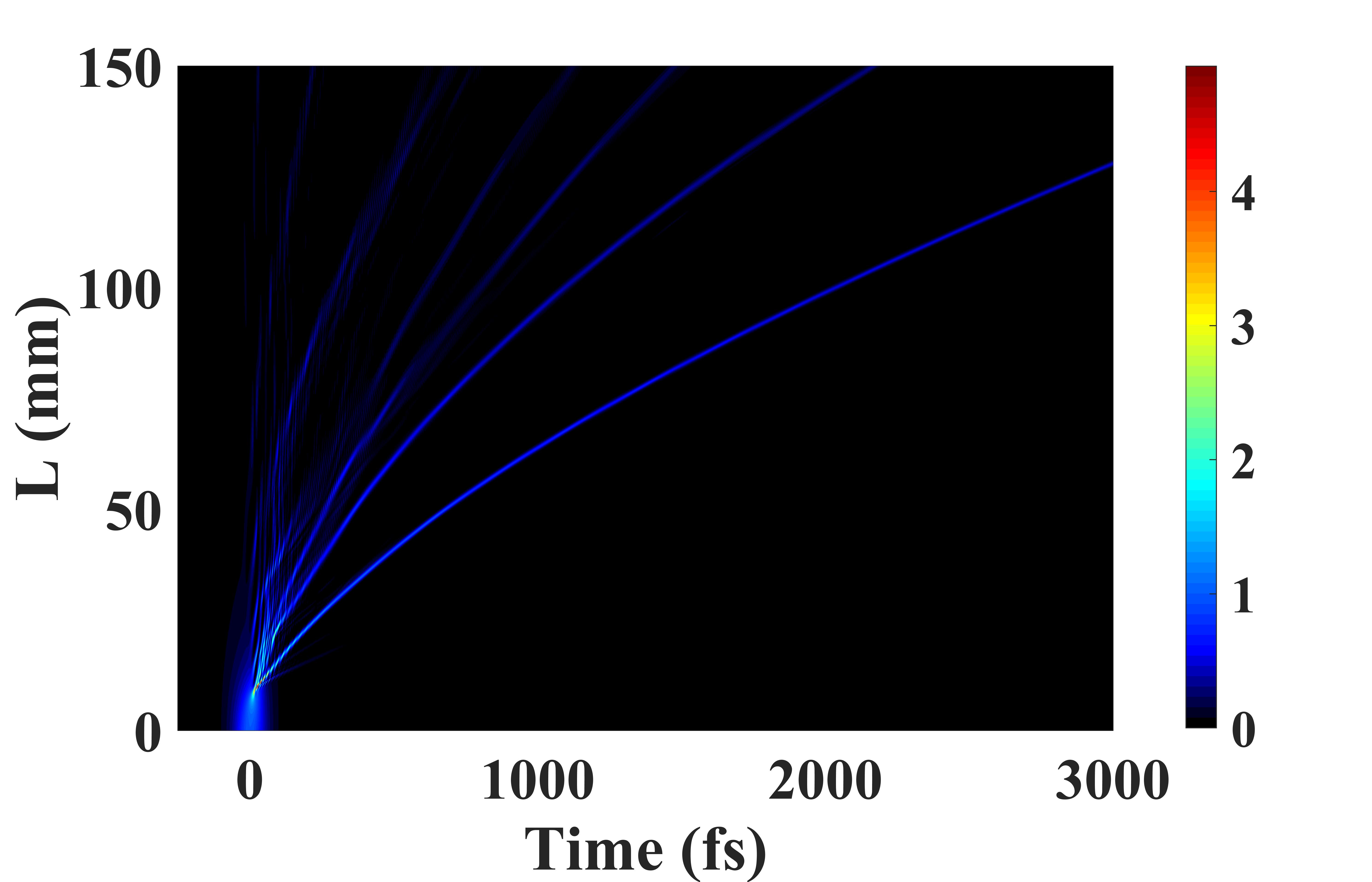}
         \caption{}
        \label{fig:1b}
    \end{subfigure}
    \begin{subfigure}[b]{0.50\textwidth}
        \includegraphics[width=\textwidth]{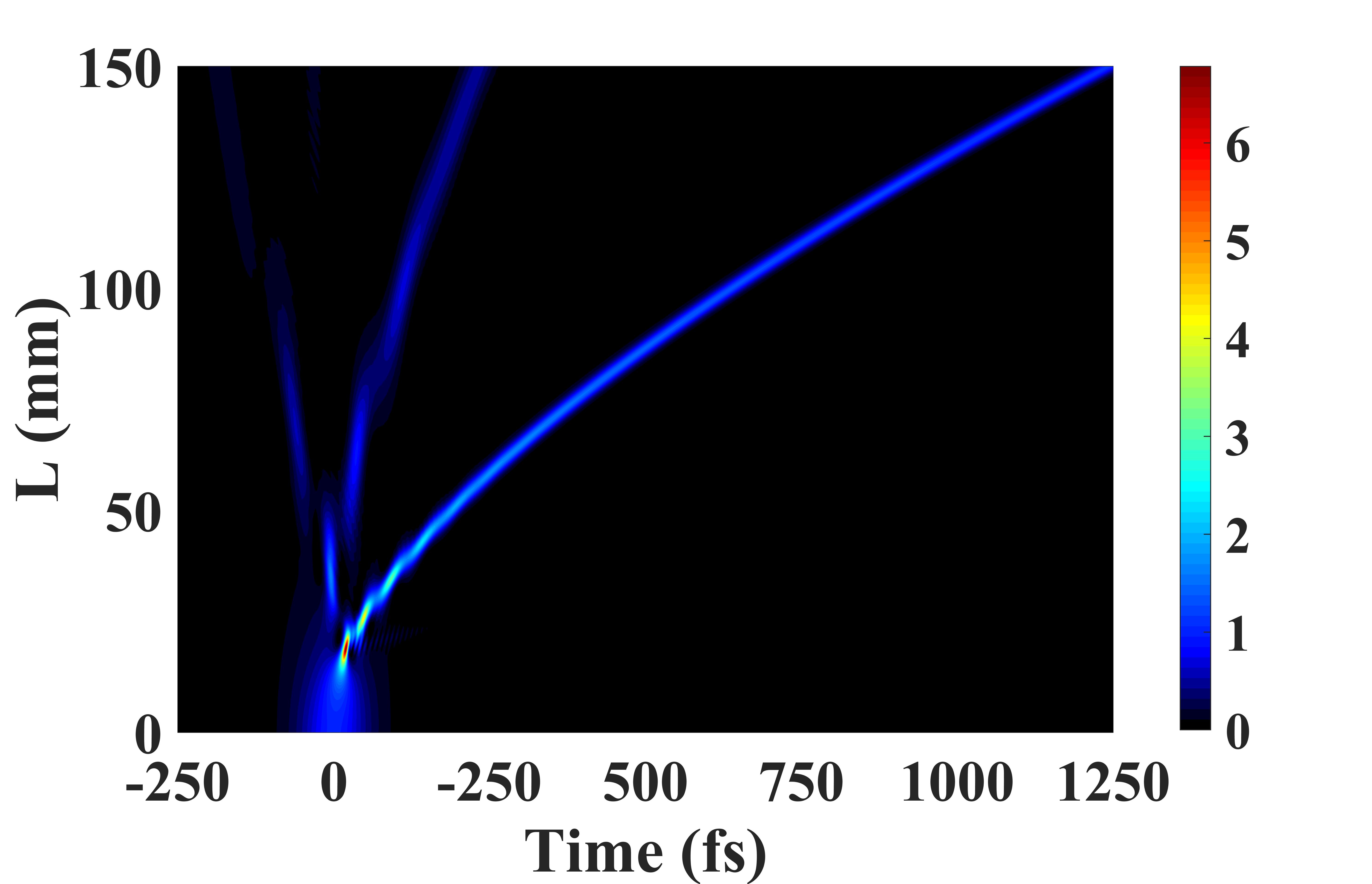}
         \caption{}
        \label{fig:1c}
    \end{subfigure}
    \begin{subfigure}[b]{0.50\textwidth}
        \includegraphics[width=\textwidth]{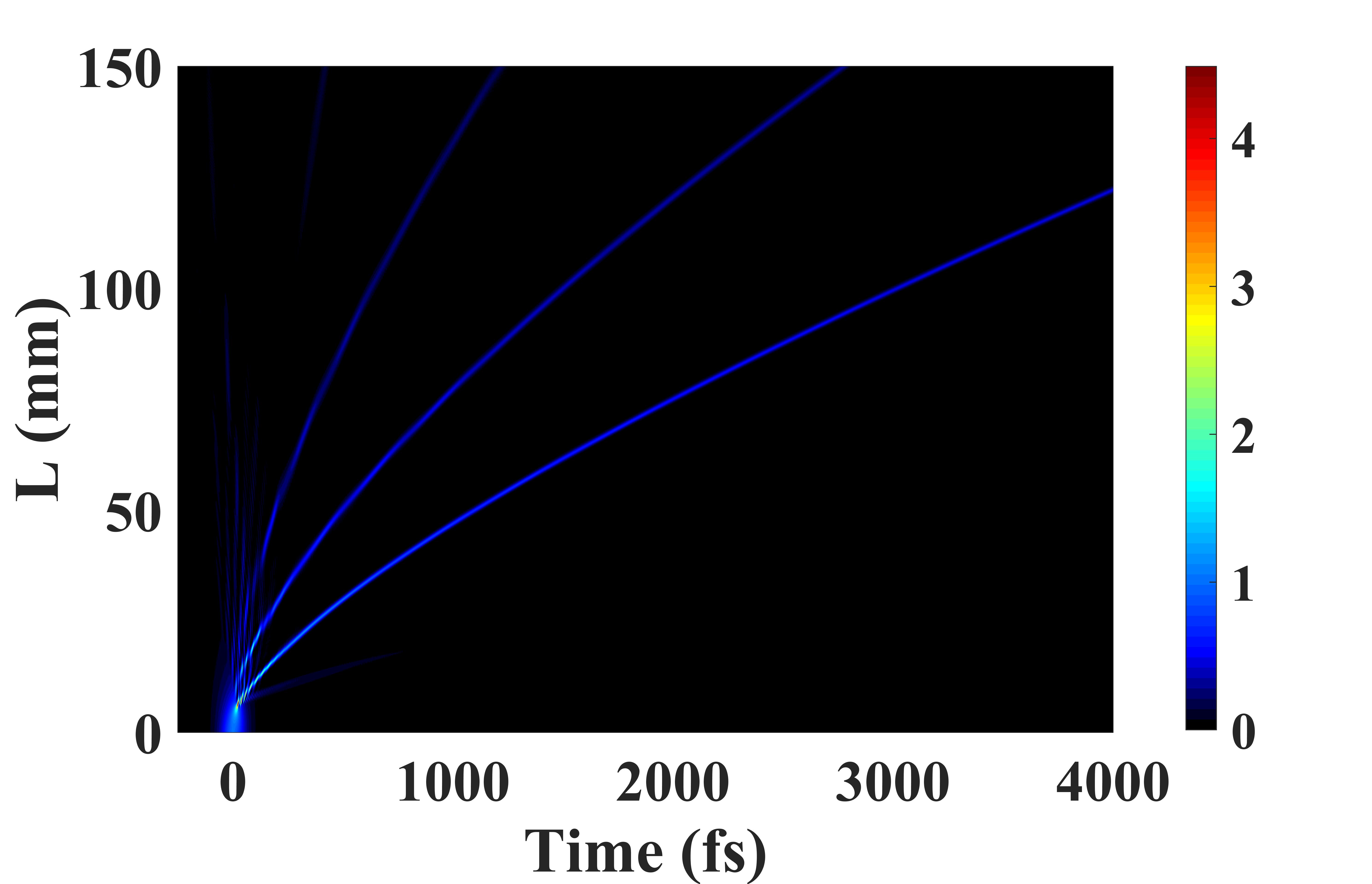}
         \caption{}
        \label{fig:1d}
    \end{subfigure}

    \caption{The temporal evolution of the high-order soliton fission produced by the (a,b) PCF, and (c,d) FSW methods. In the left column, the input peak power is $5 kW$, and in the right one it is $25 kW$.}
    \label{fig:1}
\end{figure*}

Figure 2 shows a pronounced trend for all solitons to lose their peak powers in the course of propagation over long distances. By analyzing the soliton propagation between $50 mm$ and $150 mm$, it is observed that the peak power of the first soliton is reduced by $\sim 39 \% (\sim 41 \%)$, while the second soliton experiences the decrease in the peak power of $\sim 51 \% (\sim 37 \%)$, when generated from the input power of $25 kW$, by using the PCF (FSW) method. Due to the mechanism by which multiple-soliton trains are generated, the first soliton tends to appear in the temporally isolated form, being subject to power loss via IRS due to the generation of optical phonons, while the central wavelength of the soliton features a red shift \cite{agrawal2013nonlinear}. This loss mechanism tends to limit the propagation of robust solitons over long distances, as the NL effects is weakening. Secondary solitons are also influenced by the IRS. However, as the secondary solitons are generated in the region where the energy remaining after the creation of the first soliton is concentrated, these solitons are more likely to undergo collisions with each other, giving rise to energy transfer between them. 

\begin{figure*}
    \begin{subfigure}[b]{0.55\textwidth}
        \includegraphics[width=\textwidth]{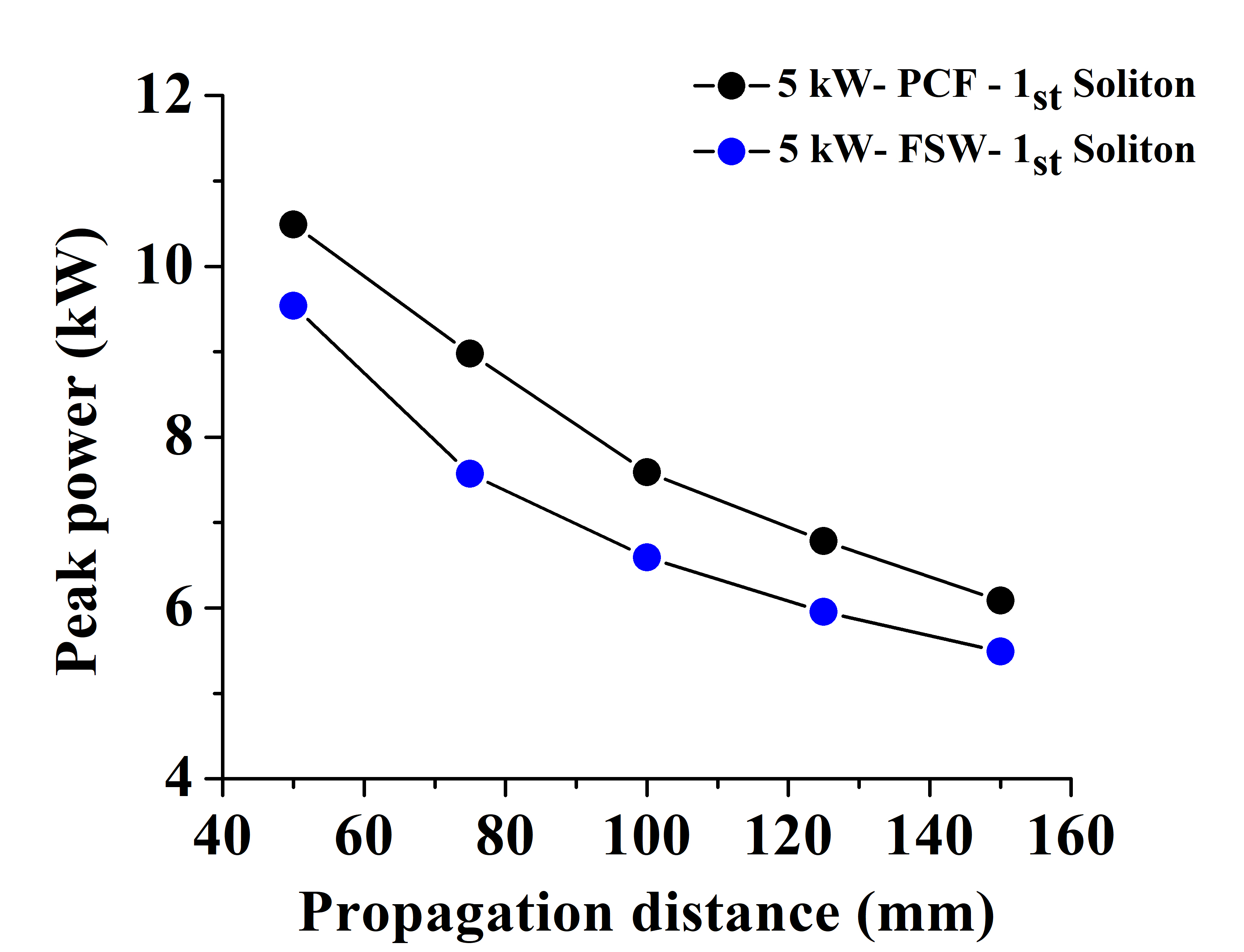}
         \caption{}
        \label{fig:2a}
    \end{subfigure}
    \begin{subfigure}[b]{0.55\textwidth}
        \includegraphics[width=\textwidth]{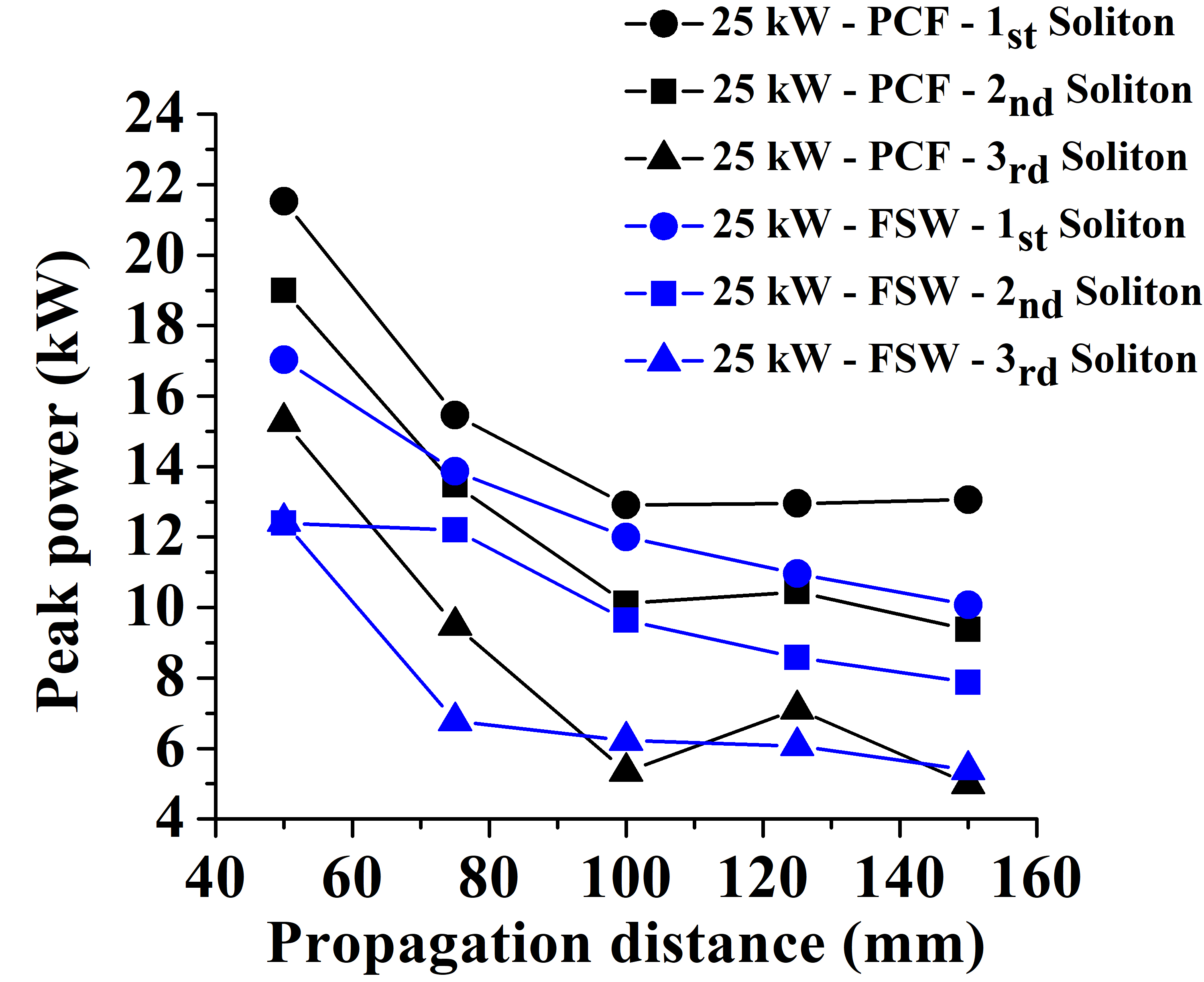}
         \caption{}
        \label{fig:2b}
    \end{subfigure}

    \caption{The evolution of the peak power of the first, second, and third solitons in the course of the propagation, in the case of using the PCF (black) or FSW (blue) method for input powers equal to (a) $5 kW$, and (b) $25 kW$.}
    \label{fig:2}
\end{figure*}

Contrary to the results produced by the PCF and FSW methods, the solution of the generalized NLSE corresponding to the MTC method provides interesting results for the generation and propagation of temporal solitons over long distances. As shown in Fig. 3, the MTC method begins by generating a pair of solitons at edges of the pulse, labeled as the leading-edge soliton ($S_{LE}$) and one at the trailing edge ($S_{TE}$). Subsequently, new pairs of secondary solitons are generated closer to the central region of the pulse. As expected, the number of generated temporal solitons increases with the increase of the input peak power. However, remarkably, the secondary solitons show powers similar to that of the first soliton, indicating more efficient energy distribution during the soliton fission. While the MTC method is efficient in generating a large number of high-power temporal solitons, we focus here on the second ($2_{nd}$) and third ($3_{rd}$) $S_{LE}$ and $S_{TE}$.

A notable difference between the three methods studied here is the dispersion regime in which solitons are generated. In the MTC method, the normal-dispersion regime provides acceleration of each soliton, as IRS causes the red shift of its central wavelength. On the contrary, the anomalous-dispersion regime in the PCF and FSW causes deceleration of the soliton in the course of propagation. This difference is the reason why the solitons in Figs. 1 and 3 propagate in opposite temporal directions. Furthermore, the soliton acceleration induced by the MTC method creates a propitious configuration to observe phenomena ranging from soliton collisions to the Newton’s cradle (NC) \cite{driben2013newton}. The latter phenomenon can be clearly seen in Fig. 3(b), where the second $S_{TE}$ collides with the first $S_{TE}$, and then collides again with the newly formed solitons in the central region of the pulse. Finally, as a result of the multiple inelastic collisions, the NC soliton ($S_{nc}$) is ejected with a higher peak power. A similar dynamical behavior, but with a larger number of collisions, is observed in the case of formation of the NC soliton at higher input powers, see Fig. 3(c).

Interestingly, in Fig.3(c) we observe the NC effect in the direction opposite to that in Fig.3(b). The central region of the pulse generates a soliton which slows down and collides, consecutively, with the third, second, and first TE solitons, and then escapes.

\begin{figure*}
    \centering
    \begin{subfigure}[b]{0.45\textwidth}
     \centering
        \includegraphics[width=\textwidth]{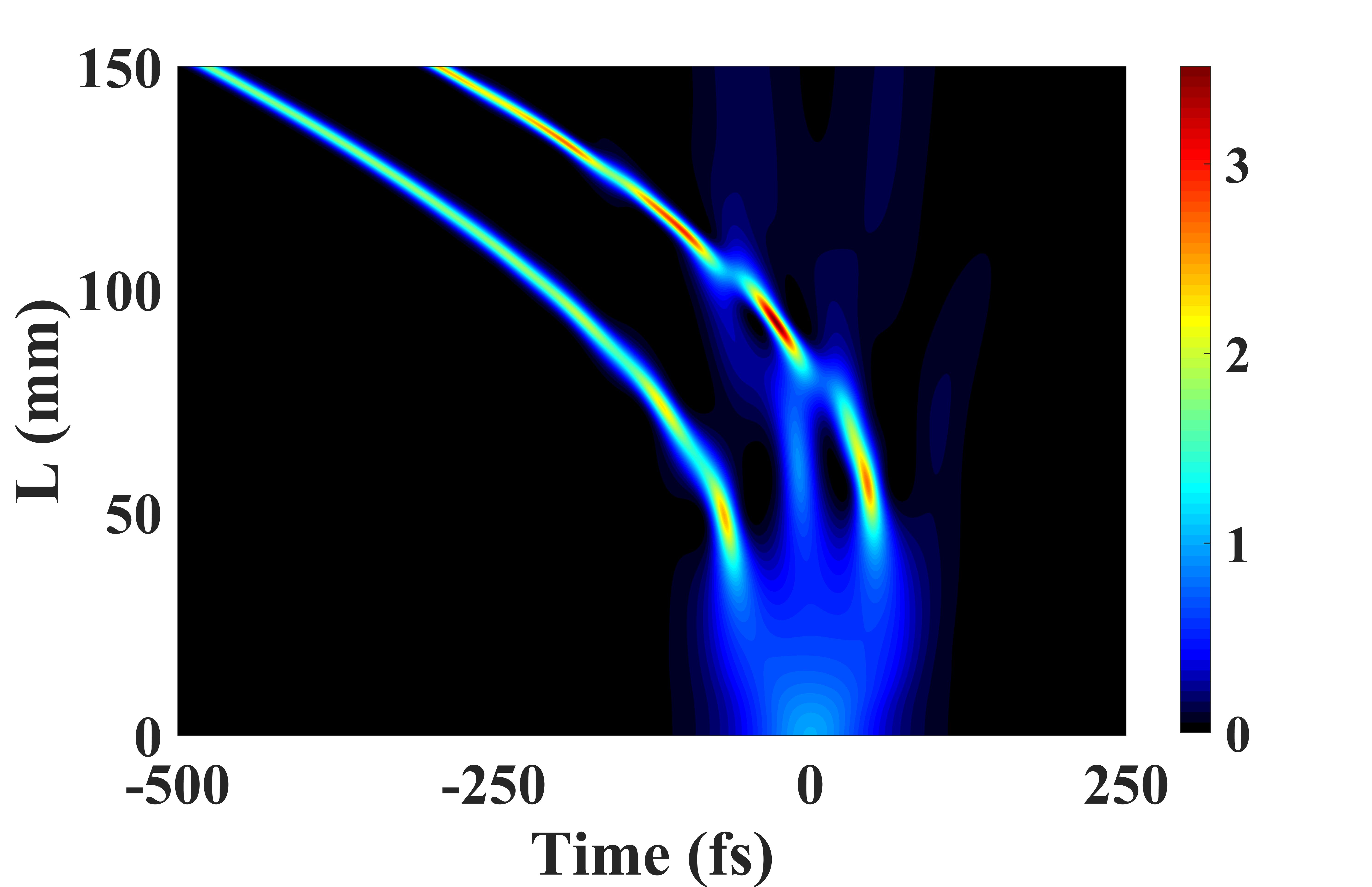}
         \caption{}
        \label{fig:3a}
    \end{subfigure}
    \begin{subfigure}[b]{0.45\textwidth}
     \centering
        \includegraphics[width=\textwidth]{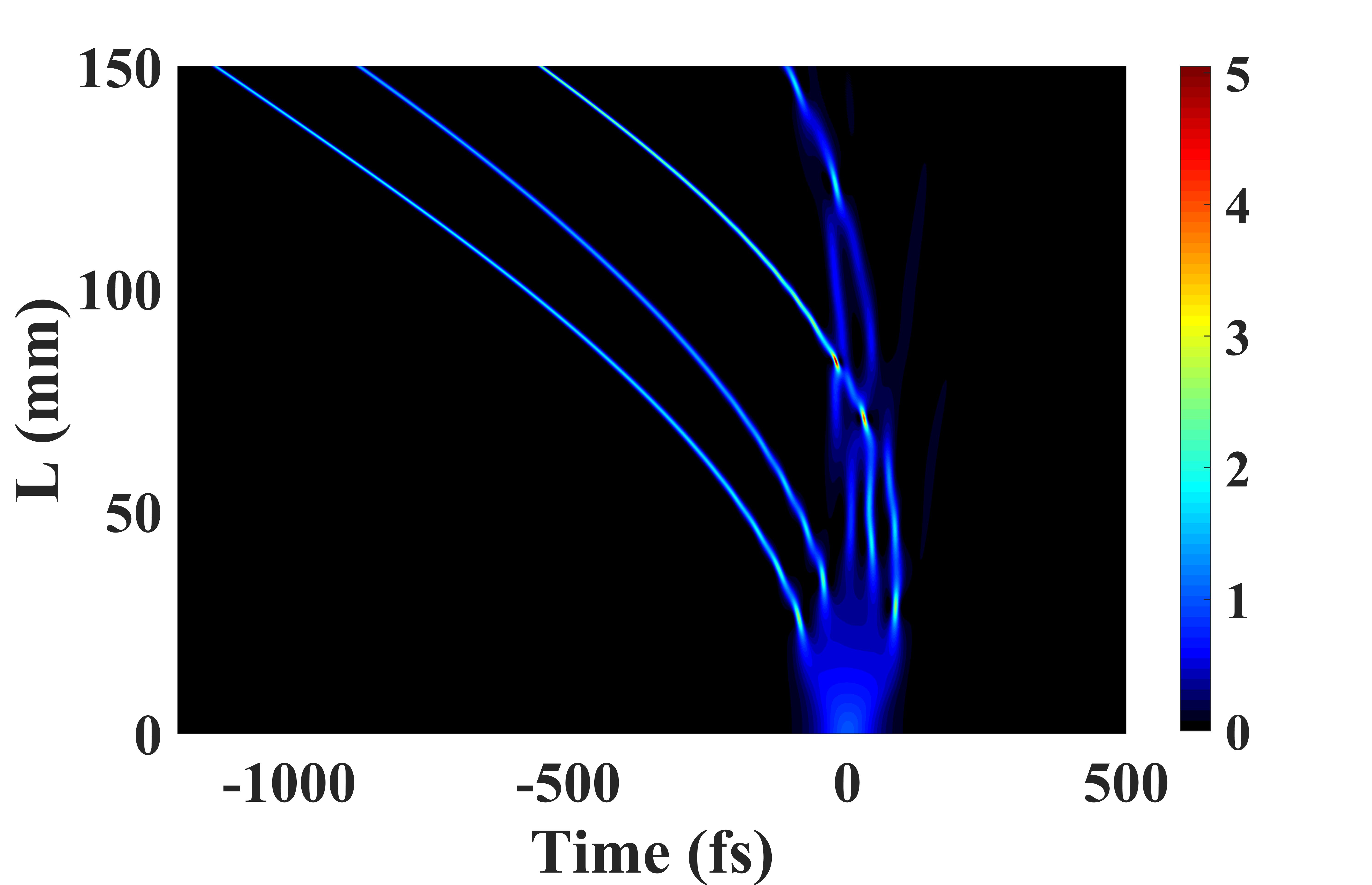}
         \caption{}
        \label{fig:3b}
    \end{subfigure}
    \begin{subfigure}[b]{0.45\textwidth}
     \centering
        \includegraphics[width=\textwidth]{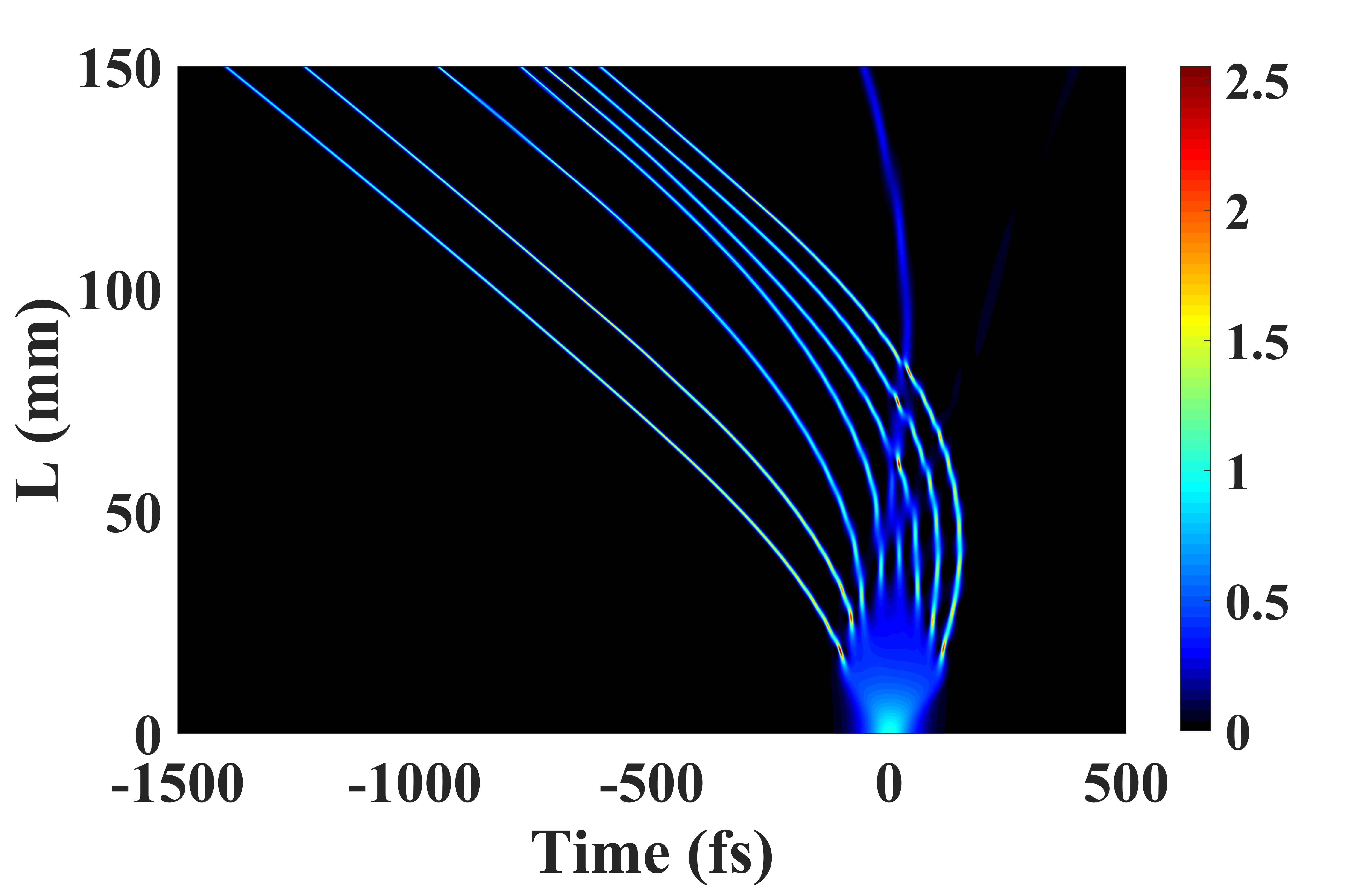}
         \caption{}
        \label{fig:3c}
    \end{subfigure}
    \begin{subfigure}[b]{0.45\textwidth}
     \centering
        \includegraphics[width=\textwidth]{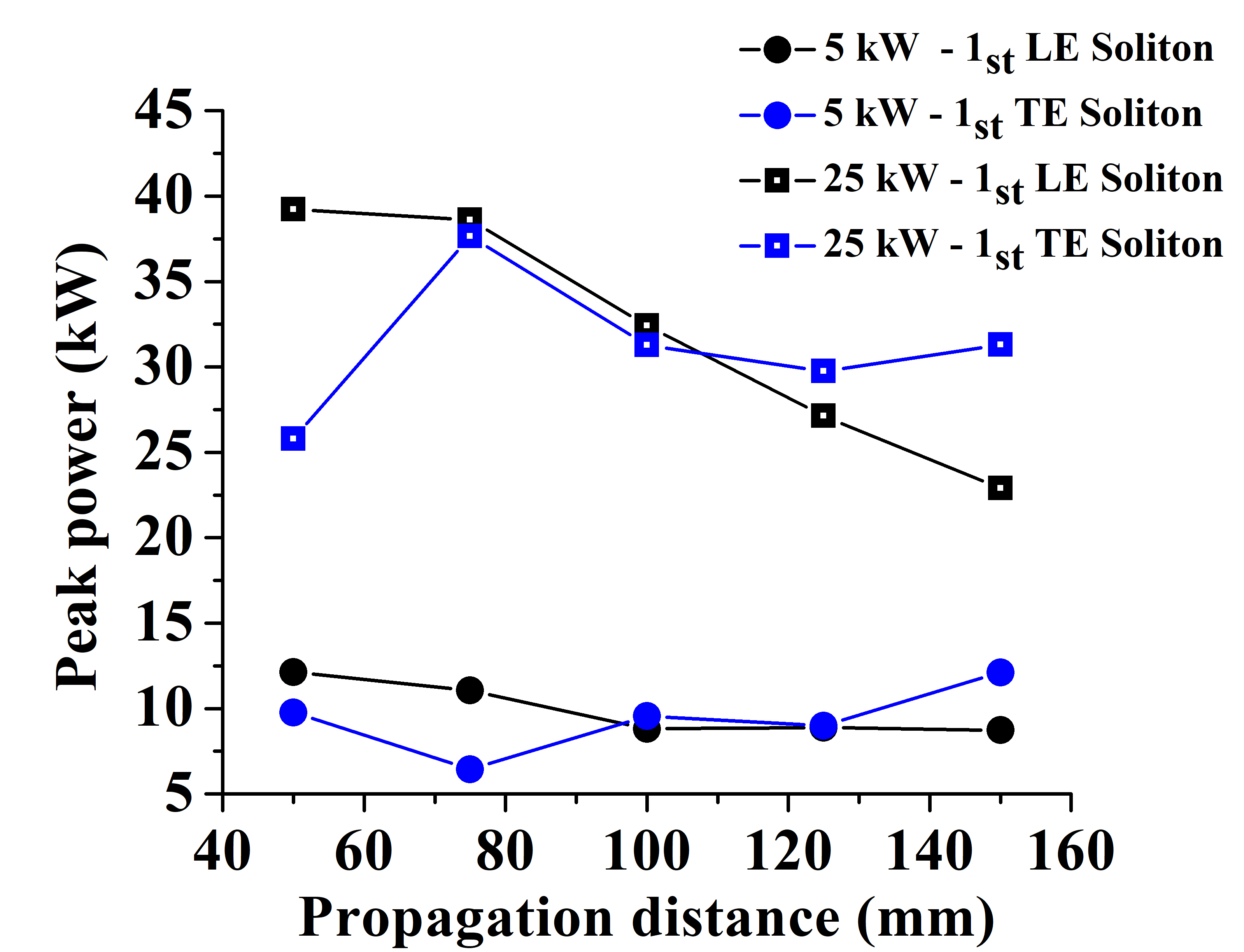}
         \caption{}
        \label{fig:3d}
    \end{subfigure}
  \begin{subfigure}[b]{0.45\textwidth}
        \includegraphics[width=\textwidth]{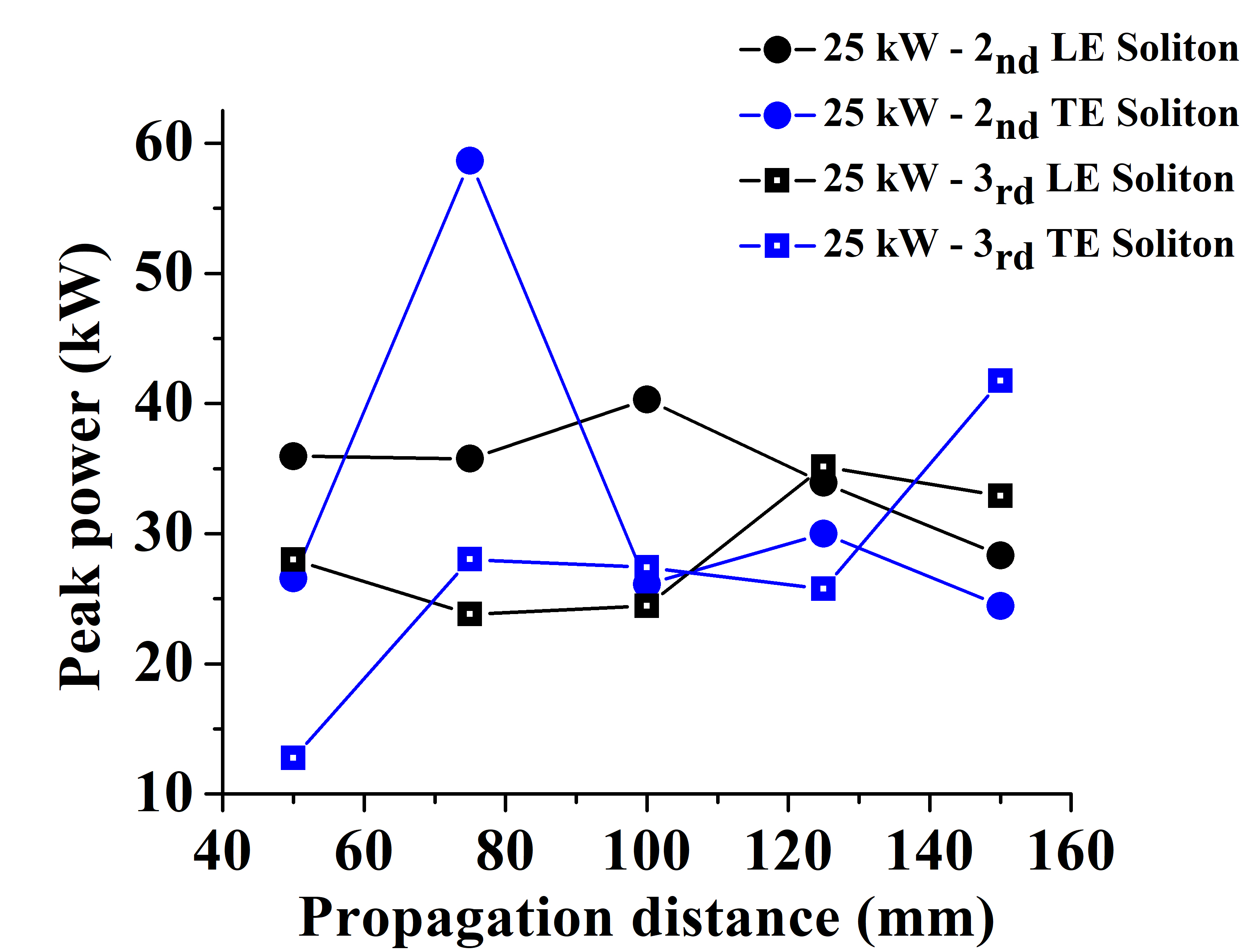}
         \caption{}
        \label{fig:1e}
    \end{subfigure}

    \caption{(a-c) The dynamics of the soliton fission produced by the MTC method, for different input powers. (d-e) The evolution of the peak powers of the first, second and third $S_{TE}$ and $S_{LE}$ generated in (a-c) in the course of the propagation.}
    \label{fig:3}
\end{figure*}

The dependence of the soliton peak power on the propagation distance, shown in Figs. 3(d,e), also shows advantages offered by the MTC method, in comparison to the PCF and FSW. In the case of the MTC method, the initial energy distribution is not only more efficient to generate multiple-soliton trains with high powers during the soliton fission, but it is also observed that each soliton has the ability to maintain its peak power during the propagation. This happens due to the cooperative process of energy exchange with dispersive waves generated during the soliton fission, or by collisions with other neighboring solitons, that allows the temporal solitons to travel longer distances compared to the case of the PCF and FSW. For example, Fig. 3(d) shows that the first $S_{TE}$ starts its propagation with a lower peak power, compared to the pair of the first $S_{LE}$. However, as the first $S_{TE}$ accelerates, the central region of the pulse transfers energy to it (see Fig. 3(a)), making its peak power comparable to that of the first $S_{LE}$. On the other hand, for higher input peak powers, the energy exchange through multiple collisions of the first $S_{TE}$ with the secondary $S_{TE}$ (Fig. 3(b)), or with the soliton NC (Fig. 3(b,c)) allows its peak power to increase or maintain its value over long propagation distances. These energy exchange and feedback processes are responsible for the fact that $S_{TE}$ demonstrates higher power than $S_{LE}$ for longer propagation distances, despite the fact that the initial energy distribution was opposite.
      
In addition, the efficiency of the PCF, FSW and MTC methods can be compared by examining the number of solitons generated with the increase of the input peak power, as shown in Fig. 4. For the same input pulse, the MTC method shows the ability to generate a larger number of temporal solitons for all peak powers, while the PCF and FSW methods exhibit poorer results. Note that, for the input peak power higher than $15 kW$, the number of solitons generated by the MTC method is almost double compared to the PCF and FSW.

\begin{figure}
    \centering
    \includegraphics [width=.60\textwidth]{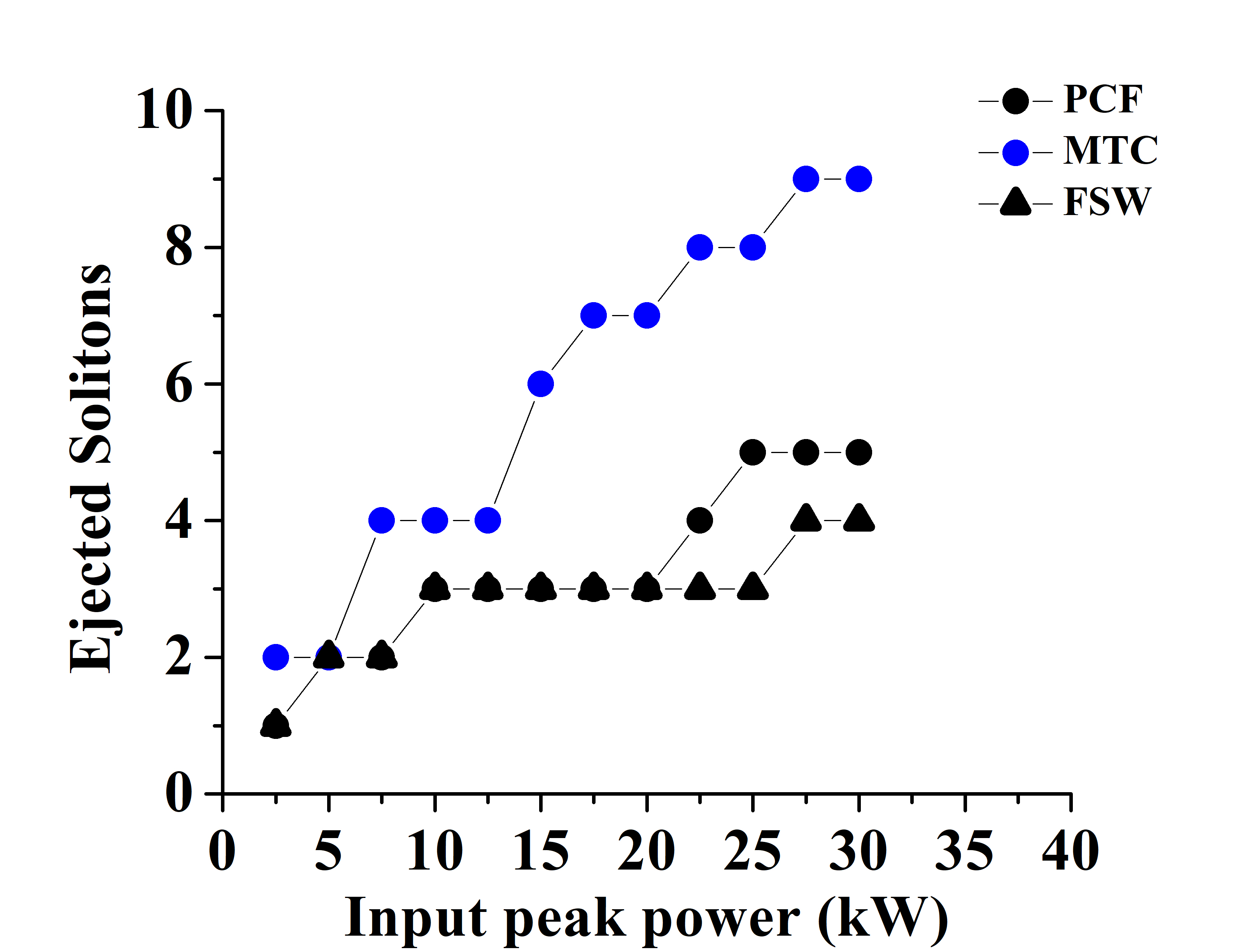}
    \caption{Numbers of solitons ejected from soliton fission as a function of the input peak power, as provided by the PCF, FSW, MTC methods.}
    \label{fig:4}
\end{figure}

Thus, the MTC method emerges as a powerful tool to generate multiple temporal solitons, with high peak power, which is highly desired for long-propagation distances. However, the efficiency of the MTC declines considerably when the goal is the generation of multiple solitons in the visible regime, as the method seems to be ineffective for generating supercontinuum spectra that reach the visible range. As shown in Fig. 5, the PCF method, widely used for soliton generation in the visible regime \cite{knight1996all, birks1997endlessly, russell2003photonic, liu2012all, dudley2002numerical, hult2007fourth, agrawal2013nonlinear, dudley2006supercontinuum}, provides a supercontinuum spectrum ranging from $480 nm$ to $1360 nm$, while the MTC and FSW are limited to the spectral region from $720 nm$ to $2600 nm$. Thus, the spectral region where the MTC method starts to work efficiently corresponds to the near-infrared, taking into regard that the IRS contribution shifts the central wavelength of the solitons up to $1200 nm$. Another interesting feature of the spectral analysis shown in Fig. 5 (c) is that the observed frequency shift in the region between $1200 nm$ and $2600 nm$ is not due to the IRS. Such giant near-infrared emission ($> 1200 nm$) seems to be correlated with the accumulation of soliton collisions, similarly to Refs. \cite{antikainen2012phase, erkintalo2010giant, erkintalo2010experimental, deng2016trapping}. Further studies are underway to better understand the present setup in this spectral regime.


\begin{figure*}
    \centering
    \begin{subfigure}[b]{0.48\textwidth}
        \includegraphics[width=\textwidth]{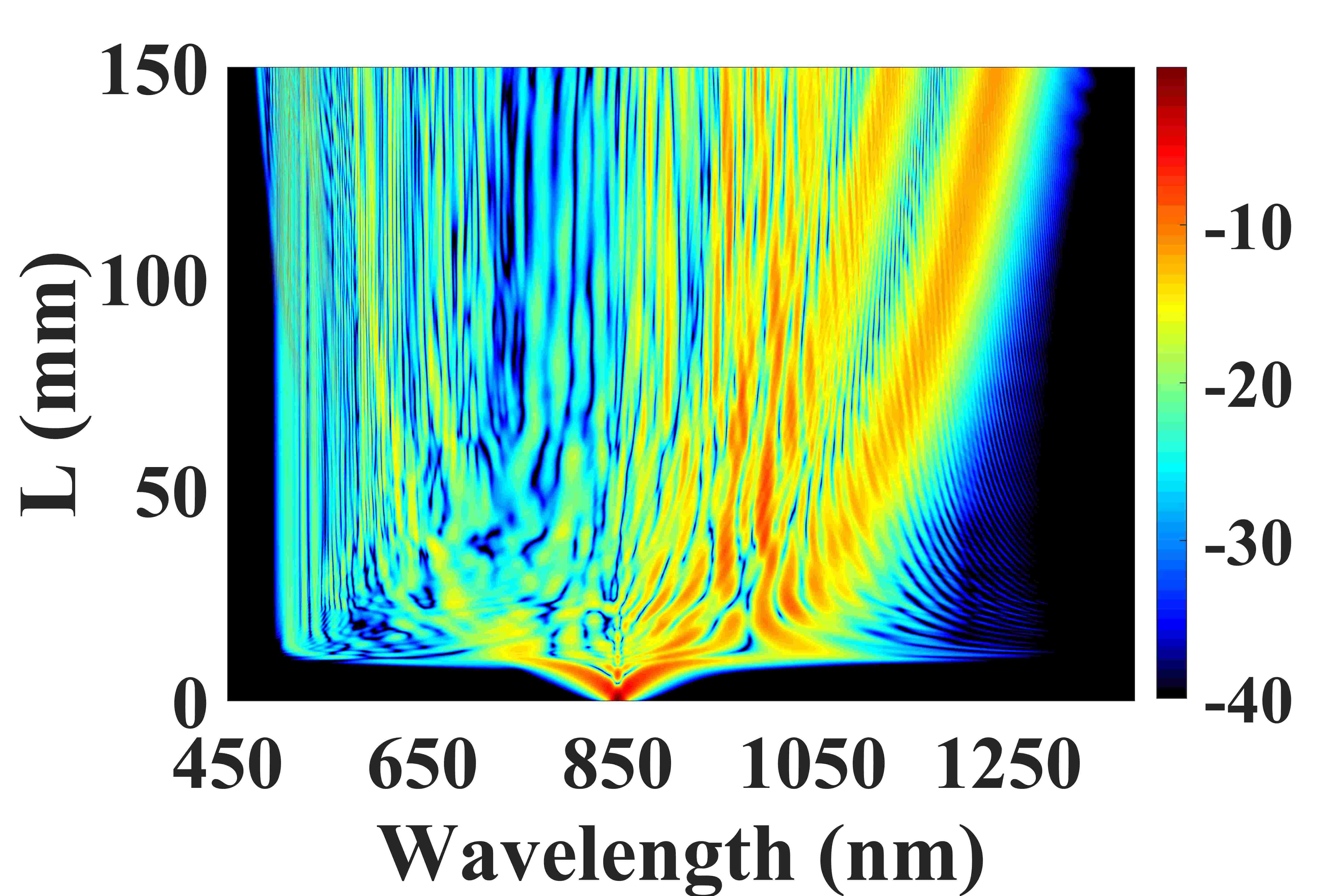}
         \caption{}
        \label{fig:5a}
    \end{subfigure}
    \begin{subfigure}[b]{0.48\textwidth}
        \includegraphics[width=\textwidth]{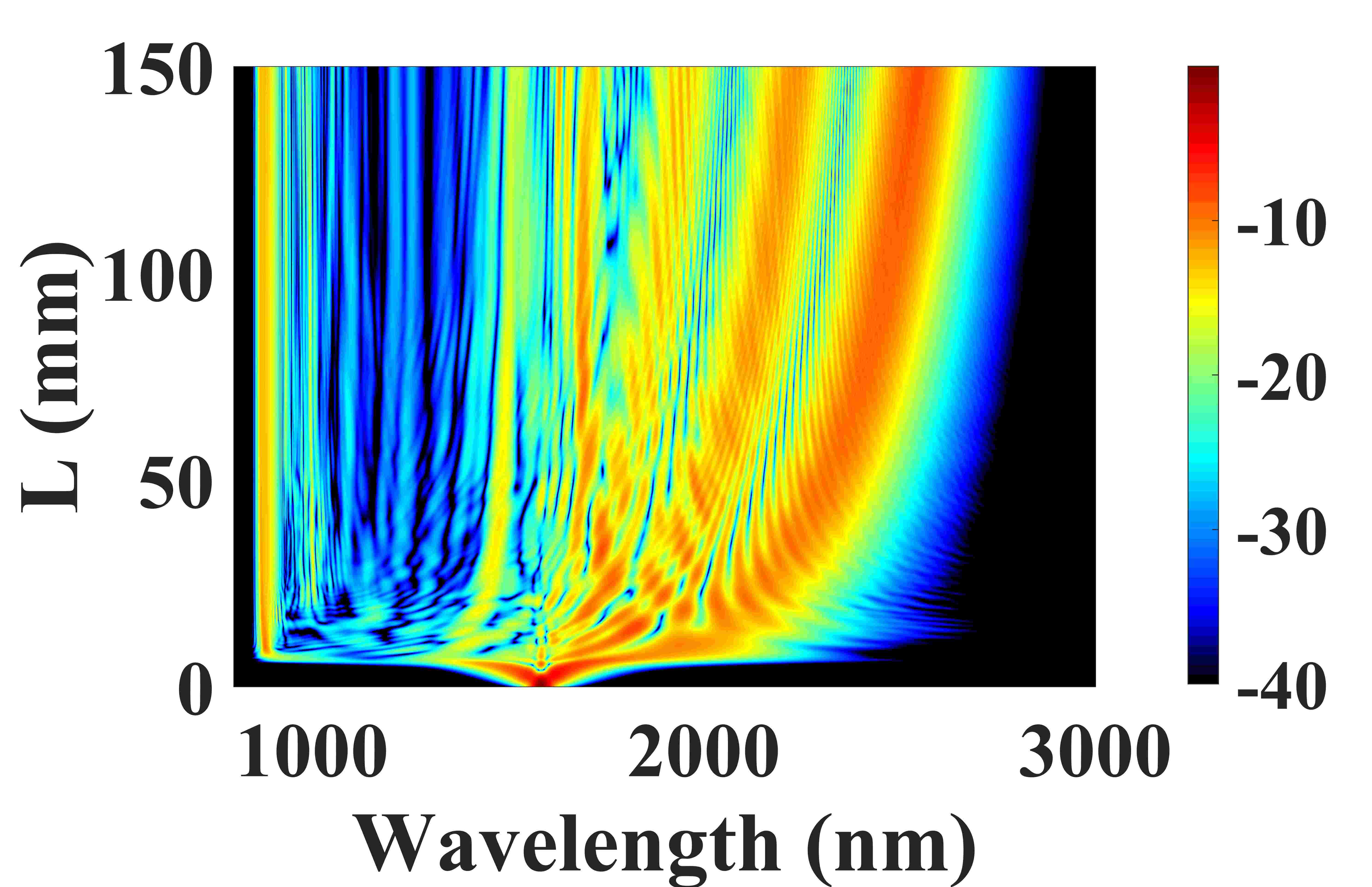}
         \caption{}
        \label{fig:5b}
    \end{subfigure}
    \begin{subfigure}[b]{0.48\textwidth}
        \includegraphics[width=\textwidth]{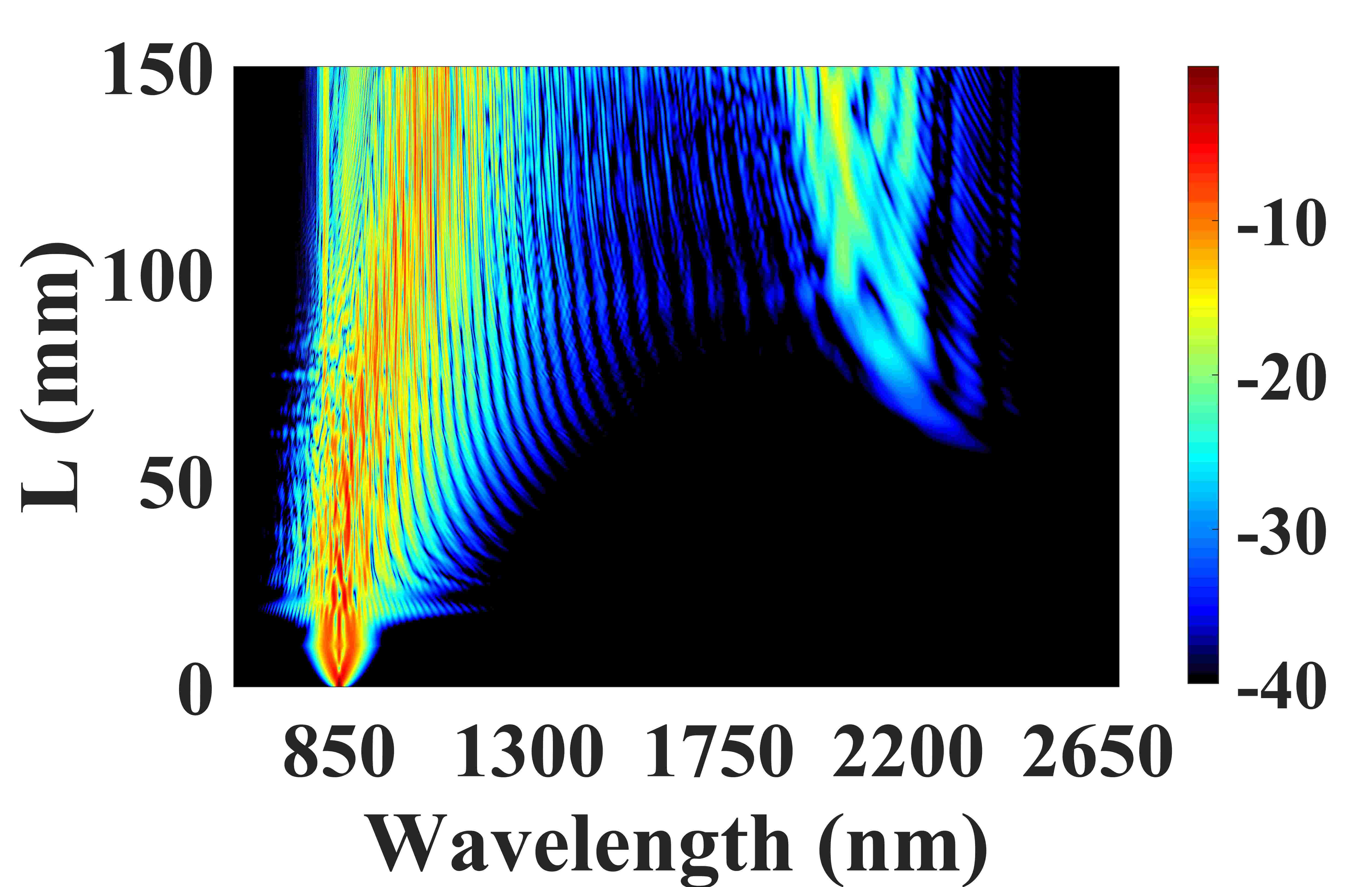}
         \caption{}
        \label{fig:5c}
    \end{subfigure}
    
    \caption{The pulse’s spectral evolution, for the input peak power $25 kW$, as produced by the (a) PCF, (b) FSW and (c) MTC methods.}
    \label{fig:5}
\end{figure*}

\section{\label{sec:level4}Conclusions}

In summary, we demonstrate, through systematic numerical calculations of the generalized nonlinear Schr\"{o}dinger equation, the high efficiency that the multiple temporal compression (MTC) method provides for the generation of multiple temporal solitons from a single optical pulse, in comparison to the well-known PCF and FSW methods. The numerical results show that the MTC method allows to generate twice as many solitons as the conventional ones, for the input peak powers studied here, with each soliton exhibiting a high power that remains approximately constant during the propagation. While the efficiency offered by the MTC technique decreases in the visible region, it appears as a useful tool for the studies of multiple soliton collisions in the near-infrared region, as well as for demonstrating the realization of the multi-soliton Newton’s cradle. 

\section{Acknowledgments} \label{sect:Acknowl}

This work was supported by the Brazilian agencies Conselho Nacional de Desenvolvimento Científico e Tecnol\'ogico - CNPq (Grant: 431162/2018-2) and the National Institute of Photonics (INCT program) - Grant: 465.763/2014, Funda\c{c}\~ao de Amparo \`a Ci\^encia e Tecnologia do Estado de Pernambuco (FACEPE), and Coordena\c c\~ao de Aperfei\c coamento de Pessoal de N\'ivel Superior (CAPES). The doctoral scholarship of A.C.A. Siqueira was provided by the CNPq. G. Palacios thanks a fellowship from CNPq Process No. 381191/2022-2. The work of B.A. Malomed was supported, in part, by the Israel Science Foundation through grant No. 1695/22.




\begin{thebibliography}{00}


\bibitem{sysoliatin2007soliton}A.A. Sysoliatin, A.K. Senatorov, A.I. Konyukhov, L.A. Melnikov, V.A. Stasyuk, Soliton fission management by dispersion oscillating fiber, Opt. Express 15 (2007) 16302-16307. https://doi.org/10.1364/OE.15.016302.
\bibitem{tai1988fission}K. Tai, A. Hasegawa, N. Bekki, Fission of optical solitons induced by stimulated Raman effect, Opt. Lett. 13 (1988) 392-394. https://doi.org/10.1364/OL.13.000392.
\bibitem{driben2013newton}R. Driben, B.A. Malomed, A.V. Yulin, D.V. Skryabin, Newton's cradles in optics: From N-soliton fission to soliton chains, Phys. Rev. A 87 (2013) 063808. https://doi.org/10.1103/PhysRevA.87.063808. 
\bibitem{braud2016solitonization}F. Braud, M. Conforti, A. Cassez, A. Mussot, A. Kudlinski,  Solitonization of a dispersive wave, Opt. Lett. 41 (2016) 1412-1415. https://doi.org/10.1364/OL.41.001412.
\bibitem{gordon1986theory}J.P. Gordon, Theory of the soliton self-frequency shift, Opt. Lett. 11 (1986) 662-664. https://doi.org/10.1364/OL.11.000662.
\bibitem{mitschke1986discovery}F.M. Mitschke, L.F. Mollenauer, Discovery of the soliton self-frequency shift, Opt. Lett. 11 (1986) 659-661. https://doi.org/10.1364/OL.11.000659.
\bibitem{xiang2011controllable}Y. Xiang, X. Dai, S. Wen, J. Guo, D. Fan, Controllable Raman soliton self-frequency shift in nonlinear metamaterials, Phys. Rev. A. 84 (2011) 033815. https://doi.org/10.1103/PhysRevA.84.033815.
\bibitem{skryabin2003soliton}D.V. Skryabin, F. Luan, J.C. Knight, P.St.J. Russell, Soliton self-frequency shift cancellation in photonic crystal fibers, Science 301 (2003) 1705-1708. https://doi.org/10.1126/science.1088516.
\bibitem{knight1996all}J.C. Knight, T.A. Birks, P.St.J. Russell, D.M. Atkin, All-silica single-mode optical fiber with photonic crystal cladding, Opt. Lett. 21 (1996) 1547-1549. https://doi.org/10.1364/OL.21.001547.
\bibitem{birks1997endlessly}T.A. Birks, J.C. Knight, P.St.J. Russell, Endlessly single-mode photonic crystal fiber, Opt. Lett. 22 (1997) 961-963. https://doi.org/10.1364/OL.22.000961.
\bibitem{russell2003photonic}P. Russell, Photonic crystal fibers, Science 299 (2003) 358-362. https://doi.org/10.1126/science.1079280.
\bibitem{liu2012all}L. Liu, Q. Tian, M. Liao, D. Zhao, G. Qin, Y. Ohishi, W. Qin, All-optical control of group velocity dispersion in tellurite photonic crystal fibers, Opt. Lett. 37 (2012) 5124-5126. https://doi.org/10.1364/OL.37.005124.
\bibitem{dudley2002numerical}J.M. Dudley, S. Coen, Numerical simulations and coherence properties of supercontinuum generation in photonic crystal and tapered optical fibers, IEEE J. Sel. Top. Quantum Electron. 8 (2002) 651-659. https://doi.org/10.1109/JSTQE.2002.1016369.
\bibitem{hult2007fourth}J. Hult, A fourth-order Runge–Kutta in the interaction picture method for simulating supercontinuum generation in optical fibers, J. Lightwave Technol. 25 (2007) 3770-3775. https://doi.org/10.1109/JLT.2007.909373. 
\bibitem{agrawal2013nonlinear}G.P. Agrawal, Nonlinear Fiber Optics, fifth ed., Academic Press, Elsevier, 2013.
\bibitem{dudley2006supercontinuum}J.M. Dudley, G. Genty, S. Coen, Supercontinuum generation in photonic crystal fiber, Rev. Mod. Phys. 78 (2006) 1135. https://doi.org/10.1103/RevModPhys.78.1135.
\bibitem{hasegawa1973transmission}A. Hasegawa, F. Tappert, Transmission of stationary nonlinear optical pulses in dispersive dielectric fibers. I. Anomalous dispersion, Appl. Phys. Lett. 23 (1973) 142-144. https://doi.org/10.1063/1.1654836. 
\bibitem{mollenauer1988demonstration}L.F. Mollenauer, K. Smith, Demonstration of soliton transmission over more than 4000 km in fiber with loss periodically compensated by Raman gain, Opt. Lett. 13 (1988) 675-677. https://doi.org/10.1364/OL.13.000675. 
\bibitem{hasegawa1995solitons}A. Hasegawa, Y. Kodama, Solitons in optical communications, Oxford University Press on Demand, 1995.
\bibitem{agrawal2012fiber}G.P. Agrawal, Fiber-optic communication systems, fourth ed., John Wiley \& Sons, 2012.
\bibitem{Iannone1998nonlinear}E. Iannone, F. Matera, A. Mecozzi, M. Settembre, Nonlinear Optical Communication Networks, Wiley, 1998.
\bibitem{agrawal2001applications}G.P. Agrawal, Applications of nonlinear fiber optics, Elsevier, 2001.
\bibitem{mollenauer2006solitons}L.F. Mollenauer, J.P. Gordon, Solitons in optical fibers: fundamentals and applications, Elsevier, 2006.
\bibitem{antikainen2012phase}A. Antikainen, M. Erkintalo, J.M. Dudley, G. Genty, On the phase-dependent manifestation of optical rogue waves, Nonlinearity 25 (2012) R73. https://doi.org/10.1088/0951-7715/25/7/R73. 
\bibitem{erkintalo2010giant}M. Erkintalo, G. Genty, J.M. Dudley, Giant dispersive wave generation through soliton collision, Opt. Lett. 35 (2010) 658-660. https://doi.org/10.1364/OL.35.000658. 
\bibitem{husakou2001supercontinuum}A.V. Husakou, J. Herrmann, Supercontinuum generation of higher-order solitons by fission in photonic crystal fibers, Phys. Rev. Lett. 87 (2001) 203901. https://doi.org/10.1103/PhysRevLett.87.203901. 
\bibitem{bose2016implications}S. Bose, A. Sahoo, R. Chattopadhyay, S. Roy, S.K. Bhadra, G.P. Agrawal, Implications of a zero-nonlinearity wavelength in photonic crystal fibers doped with silver nanoparticles, Phys. Rev. A 94 (2016) 043835. https://doi.org/10.1103/PhysRevA.94.043835. 
\bibitem{arteaga2018soliton}F.R. Arteaga-Sierra, A. Antikainen, G.P. Agrawal, Soliton dynamics in photonic-crystal fibers with frequency-dependent Kerr nonlinearity, Phys. Rev. A 98 (2018) 013830. https://doi.org/10.1103/PhysRevA.98.013830.
\bibitem{bose2016study}S. Bose, R. Chattopadhyay, S. Roy, S. K. Bhadra, Study of nonlinear dynamics in silver-nanoparticle-doped photonic crystal fiber,  J. Opt. Soc. Am. B 33 (2016) 1014-1021. https://doi.org/10.1364/JOSAB.33.001014.  
\bibitem{bose2018dispersive}S. Bose, R. Chattopadhyay, S.K. Bhadra, Dispersive shock mediated resonant radiations in defocused nonlinear medium, Opt. Commun. 412 (2018) 226-229. https://doi.org/10.1016/j.optcom.2017.12.016. 
\bibitem{driben2010solitary}R. Driben, J. Herrmann, Solitary pulse propagation and soliton-induced supercontinuum generation in silica glasses containing silver nanoparticles, Opt. Lett. 35 (2010) 2529-2531. https://doi.org/10.1364/OL.35.002529.
\bibitem{zhao2022effects}S. Zhao, R. Guo, Y. Zeng, Effects of frequency-dependent Kerr nonlinearity on higher-order soliton evolution in a photonic crystal fiber with one zero-dispersion wavelength, Phys. Rev. A 106 (2022) 033516. https://doi.org/10.1103/PhysRevA.106.033516. 
\bibitem{skryabin2010colloquium}Skryabin, D. \& Gorbach, A. Colloquium: Looking at a soliton through the prism of optical supercontinuum. Rev. Mod. Phys. 82 (2010) 1287. https://doi.org/10.1103/RevModPhys.82.1287.
\bibitem{siqueira2023generation}A.C.A. Siqueira, E.L Falc\~ao-Filho, B.A Malomed, \& C.B de Ara\'ujo. Generation of multiple ultrashort temporal solitons in a third-order nonlinear composite medium with self-focusing and self-defocusing nonlinearities. {\em Physical Review A}. \textbf{107}, 063519 (2023).
https://doi.org/10.1103/PhysRevA.107.063519
\bibitem{siqueira2022generation}A.C.A. Siqueira, B.A. Malomed, C.B. de Ara\'ujo, E.L. Falc\~ao-Filho, Generation of multiple ultrashort solitons in heterogeneous medium with self-focusing and defocusing nonlinearities, Latin America Optics And Photonics Conference, pp. M4A-3 (2022). https://doi.org/10.1364/LAOP.2022.M4A.3. 
\bibitem{malomed2006soliton}B.A. Malomed, Soliton management in periodic systems, Springer, 2006. https://doi.org/10.1007/0-387-29334-5
\bibitem{malitson1965interspecimen}I.H. Malitson, Interspecimen comparison of the refractive index of fused silica, J. Opt. Soc. Am. 55 (1965) 1205-1209. https://doi.org/10.1364/JOSA.55.001205.
\bibitem{zhang2017nonlinear}Y. Zhang, Y. Wang, Nonlinear optical properties of metal nanoparticles: a review, RSC Adv. 7 (2017) 45129-45144. https://doi.org/10.1039/C7RA07551K.
\bibitem{reyna2017high}A.S. Reyna, C.B. de Ara\'ujo, High-order optical nonlinearities in plasmonic nanocomposites—a review, Adv. Opt. Phot. 9 (2017) 720-774. https://doi.org/10.1364/AOP.9.000720.
\bibitem{kassab2018metal}L.R.P. Kassab, C.B. de Ara\'ujo, Metal nanostructures for photonics, Elsevier, 2018.
\bibitem{reyna2022beyond}A.S. Reyna, C.B. de Araújo, Beyond third-order optical nonlinearities in liquid suspensions of metal-nanoparticles and metal-nanoclusters. J. Opt. 24 (2022) 104006. https://doi.org/10.1088/2040-8986/ac8b94. 
\bibitem{desalvo1992self}R. DeSalvo, D.J. Hagan, M. Sheik-Bahae, G. Stegeman, E.W.V. Stryland, H. Vanherzeele, Self-focusing and self-defocusing by cascaded second-order effects in KTP, Opt. Lett. 17 (1992) 28-30. https://doi.org/10.1364/OL.17.000028.
\bibitem{ashihara2002soliton}S. Ashihara, J. Nishina, T. Shimura, K. Kuroda, Soliton compression of femtosecond pulses in quadratic media, J. Opt. Soc. Am. B 19 (2002) 2505-2510. https://doi.org/10.1364/JOSAB.19.002505. 
\bibitem{bache2008limits}M. Bache, O. Bang, W. Krolikowski, J. Moses, F.W. Wise, Limits to compression with cascaded quadratic soliton compressors. Opt. Express 16 (2008) 3273-3287. https://doi.org/10.1364/OE.16.003273.
\bibitem{guo2014few}H. Guo, X. Zeng, B. Zhou, M. Bache, Few-cycle solitons and supercontinuum generation with cascaded quadratic nonlinearities in unpoled lithium niobate ridge waveguides, Opt. Lett. 39 (2014) 1105-1108. https://doi.org/10.1364/OL.39.001105.
\bibitem{vsuminas2017second}R. \v{S}uminas, G. Tamo\v{s}auskas, V. Jukna, A. Couairon, A. Dubietis, Second-order cascading-assisted filamentation and controllable supercontinuum generation in birefringent crystals, Opt. Express 25 (2017) 6746-6756. https://doi.org/10.1364/OE.25.006746. 
\bibitem{conforti2013extreme}M. Conforti, F. Baronio, Extreme high-intensity and ultrabroadband interactions in anisotropic $\beta$-BaB$_2$O$_4$ crystals, J. Opt. Soc. Am. B 30 (2013) 1041-1047. https://doi.org/10.1364/JOSAB.30.001041.
\bibitem{erkintalo2010experimental}M. Erkintalo, G. Genty, J.M. Dudley, Experimental signatures of dispersive waves emitted during soliton collisions, Opt. Express 18 (2010) 13379-13384. https://doi.org/10.1364/OE.18.013379.
\bibitem{deng2016trapping}Z. Deng, X. Fu, J. Liu, C. Zhao, S. Wen, Trapping and controlling the dispersive wave within a solitonic well, Opt. Express 24 (2016) 10302-10312. https://doi.org/10.1364/OE.24.010302

\end{thebibliography}


\end{document}